\documentclass[reqno,11pt]{amsart}
\usepackage{amsmath,amssymb,tabularx,setspace,color,multirow}
\usepackage[a4paper, total={6in, 8in}]{geometry}
\usepackage[utf8]{inputenc}
 \usepackage[T1]{fontenc}
 \usepackage{mathtools}
 \usepackage{breqn}
\newtheorem{theorem}{Theorem}

\newtheorem{definition}{Definition}
\newtheorem{remark}{Remark}

\makeatletter
\renewcommand\section{\@startsection {section}{1}{\z@}%
                                   {-3.5ex \@plus -1ex \@minus -.2ex}%
                                   {2.3ex \@plus.2ex}%
                                   {\normalfont\large\bfseries}}
\makeatother

\newtheorem{Theorem}{Theorem}

\newtheorem{Example}{Example}

\begin{document}
\doublespace
\title[]{ JEL ratio test for independence of time to failure and cause of failure in competing risks}
\author[]%
{   S\lowercase{reelakshmy} N.\lowercase{\textsuperscript{a} and } S\lowercase{reedevi} E. P.\textsuperscript{\lowercase{b}}\\
 \lowercase{\textsuperscript{a}}P\lowercase{rajyothi} N\lowercase{ikethan} C\lowercase{ollege},
  P\lowercase{udukkad}, I\lowercase{ndia.}\\ \lowercase{\textsuperscript{b}}SNGS C\lowercase{ollege}, P\lowercase{attambi}, I\lowercase{ndia}.}
\thanks{{$^{\dag}$}{Corresponding E-mail: \tt sreedeviep@gmail.com}}

\begin{abstract}
In the present article,  we propose  jackknife empirical likelihood (JEL) ratio test for testing the independence of time to failure and cause of failure in competing risks data. We use U-statistic theory to derive the  JEL ratio test.  The asymptotic distribution of the test statistic is shown to be chi-square distribution with one degree of freedom. A Monte Carlo simulation study is carried out to assess the finite sample behaviour of the proposed test. The performance of proposed JEL test is compared with the test given in Dewan et al. (2004). Finally we illustrate our test procedure using various real data sets.
\end{abstract}
\maketitle

\section{Introduction}
\par
In survival studies, often the failure of individuals may be attributed to more than one cause of failure. For example, with human beings, the primary cause of death may be classified as cancer, heart disease or other causes. Competing risks models are used to analyse such situations. In competing risk analysis, we need to estimate the marginal probability of the occurrence of  a certain event when  the competing events are present.  In such cases, traditional methods of survival analysis like Kaplan-Meier (Product-limit) method can not be applied. Here we use the concept of cumulative incidence functions or cause specific hazard rate functions to analyse the marginal probability of cause-specific events.

 \par Consider a general competing risks set up with $k$ possible causes of failure.  Competing risks data can be represented  as a bivariate random pair $(T, J)$ where $T$ is the failure time of a subject and $J \in \{1,2,...,k\}$ is the corresponding cause of failure. Now the joint distribution of $(T,J)$, i.e. sub distribution functions are defined as
$$F_r(t)=P(T\le t,J=r),\quad r=1,2,...,k.$$
The over all distribution function of $T$ is given by $ F(t)=P(T\le t)=\sum\limits_{r=1}^{k}F_r(t).$
The cause specific hazard rate functions which give the instantaneous rate of failure due to cause $r$ is specified by
\begin{equation*}
   \lambda_r(t)=\frac{f_r(t)}{\bar{F}(t)}, \quad r=1,2,...,k,
\end{equation*}
where $f_{r}(t)$ is the cause specific density function and $\bar{F}(t)=1-F(t)$ is the survival function of $T$. For more details on competing risks data analysis, one can refer to Prentice et al. (1978), Kalbfleisch and Prentice (2002), Lawless (2011) and Crowder (2012) among others.

\par In the literature, competing risks data are either analysed  through a latent failure time approach or by considering  it as a bivariate random pair $(T,J)$. The approach based on the observable random pair $(T,J)$ helps to overcome the identifiability issue that may arise with the latent failure time approach. In modelling of competing risks data using observable random vector $(T,J)$ the nature of dependence between $T$ and $J$ is very important. If $T$ and $J$ are independent, then $F_r(t)= P(J = r)F(t)$ and hence one can study $T$ and $J$ separately (Anjana et al., 2019). Furthermore, the time to failure and cause of failure are independent if and only if the cause-specific hazard functions are proportional (Crowder, 2012). Many authors studied the tests for the independence of $T$ and $J$  including Dykstra et al. (1998), Dewan et al. (2004), Dewan et al. (2013), Sankaran et al. (2017) and Anjana et al. (2019) and references therein.
\par Empirical likelihood (EL) is a non-parametric inference tool firstly used by Thomas and Grunkemeier (1975). The general methodology is developed in the pioneer papers by Owen (1988, 1990).  This approach enjoys the wide acceptance among the researchers as it combines the effectiveness of the likelihood approach with the reliability of non-parametric procedure.  Empirical  likelihood finds applications in regression, econometrics and survival analysis. For more details on EL based works on survival analysis, one can refer to Wang and Jing (2001), Li and Wang (2003), Zhou (2015), Huang and Zhao (2018), Yu and Zhao (2019). Recently Variath and Sankaran (2020) developed a test to compare cumulative incidence functions of competing risks data using empirical likelihood approach.

\par  However, in empirical likelihood approach,  we need to maximize the non-parametric likelihood function subject to some constraints. When the constraints are non linear, it is difficult to apply EL procedure.  Thus Jing et al. (2009) introduced the jackknife empirical likelihood (JEL) approach which combines the  two popular non-parametric approaches, the jackknife method and the empirical likelihood approach. In spite of the technical feasibility and lucidity of JEL method, it is less explored in competing risks analysis. This motivated us to revisit the problem of testing independence of time to failure and cause of failure  and propose a new U-statistic based JEL ratio test statistic. To the best of our knowledge, this is the first attempt to employ JEL ratio test methodology in competing risks analysis.

\par Rest of the paper is organized as follows. In Section 2, we develop  JEL based test for testing the independence between time to failure $T$ and cause of failure $J$. We prove that the JEL ratio test statistic is asymptotically distributed as chi-square distribution with one degree of freedom.  A Monte Carlo simulation study is carried out in Section 3 to assess the finite sample  performance of the proposed test. The  test procedure is illustrated by applying it to various real data sets and the results are reported in section 4.  Finally, Section 5, summarizes major conclusions of the study.


\vspace{-0.3cm}
\section{Test Statistic}
In this study, we consider the situation with two causes, $k=2$. In the analysis of competing risks, often the interest is focused on a particular event type, where the events due to all other causes can be combined into one. Hence the above assumption  have not much impact in the study. Let $(T_i, J_i)$ for $i=1,...,n$ be  $n$  independent and identically distributed observations from $(T,J)$.  Our interest is to test the null  hypothesis
\begin{equation}
\label{2.1}
H_0: T \,\text{and}\, J\, \text{are independent}\nonumber
\end{equation}
against the alternative hypothesis
\begin{equation*}
  H_1: T\, \text{and}\, J\, \text{are not independent}.
\end{equation*}
To develop the test, we define the sub survival functions as $S_r(t)=P(T>t,J=r),~~\quad r=1,2.$, where the over all survival function of $T$ is given by $S(t)=P(T>t)=S_1(t)+S_2(t).$
Also define conditional probabilities
 \begin{equation}\label{cp}
  \phi_r(t)=P(J=r|T>t),\quad r=1,2.
\end{equation}
Note that $T$ and $J$ are independent if and only if $\phi_1(t)$ is a constant (Dewan et al., 2004). Hence $H_0$ can be written as
 \begin{equation*}
 H_0: \phi_1(t)\,\,\, \text{is a constant}\nonumber
\end{equation*}
against the alternative hypothesis
\begin{equation*}
  H_1: \phi_1(t)\, \,\,\text{is non-decreasing}.
\end{equation*}
Now the conditional probability $ \phi_1(t)$ can be written as
\begin{equation*}
  \phi_1(t)=\frac{S_1(t)}{S(t)}=\frac{1}{1+\frac{S_2(t)}{S_1(t)}}.
\end{equation*}
Hence under $H_1$, $\frac{S_2(t)}{S_1(t)}$ is non-increasing function of $t$. To construct the test, we propose a measure of deviation from $H_0$ towards $H_1$ as
\begin{equation*}\label{}
 \delta=S_1(t)f_2(t)-S_2(t)f_1(t).
 \end{equation*}
 Clearly $\delta$ is zero under $H_0$ and positive under $H_1$. Larger values of $\delta$ imply a clear departure from $H_0$ towards $H_1$. Thus we develop the test based on the departure measure
 \begin{equation}\label{depm}
  \Delta=\int_{0}^{\infty}(S_1(t)f_2(t)-S_2(t)f_1(t))dt.
\end{equation}
To construct the JEL ratio test, first we find a  U-statistic based estimator of $\Delta$. After simplification, we can rewrite $\Delta$ as
\begin{equation*}\label{eq3}
  \Delta=P(T_1>T_2,J_1=1,J_2=2)- P(T_1>T_2,J_1=2,J_2=1).
\end{equation*}
Define the kernel
\begin{equation*}
\label{2.4}
\psi^*((T_1,J_1),(T_2,J_2))=\begin{cases}
1\, &\text{if} \,\, T_1>T_2,J_1=1,J_2=2\\
-1&\, \text{if} \,\, T_1>T_2,J_1=2,J_2=2\\
0 & \text{otherwise}.
\end{cases}
\end{equation*} 
Clearly $E(\psi^*((T_1,J_1),(T_2,J_2)))=\Delta$. Let $\psi((T_1,J_1),(T_2,J_2))$ be the symmetric version of the kernel $\psi^*((T_1,J_1),(T_2,J_2))$. Therefore an unbiased  estimator of $\Delta$  is given by
\begin{equation}\label{est}
\widehat\Delta=\frac{2}{n(n-1)}  \sum_{i=1}^{n}\sum_{l=1,l<i}^{n}\psi{((T_i,J_i),(T_l,J_l))}.
\end{equation}
Now $\widehat\Delta$ is a consistent estimator of $\Delta$ (Lehmann, 1951). Next we find the asymptotic distribution of $\widehat\Delta$.
\begin{theorem}
  As $n \to \infty$, the distribution of $\sqrt{n}\left(\widehat\Delta-\Delta\right)$ is Gaussian with mean 0 and variance $\sigma^{2}$ where $4\sigma^{2}$ is given by
   $$\sigma^{2}=Var\left[E(\psi((T_1,J_1),(T_2,J_2))|T_1,J_1)\right].$$
\end{theorem}
The proof follows from the central limit theorem for U-statistics (Lee, 2019). 

\noindent Finding the asymptotic null variance is not easy. In such a a situation, the implementation of normal based test is not advisable. Also, when we have non-linear constraints in the optimization problem, implementation of empirical likelihood theory is very difficult. Hence we develop a jackknife empirical likelihood (JEL) ratio test  for testing the independence between cause of failure and time to failure.
\par Next we  derive the jackknife empirical likelihood ratio test based on  $\Delta$.  The  jackknife pseudo-values for $\Delta$ are given by
\begin{equation}\label{jsv}
\widehat{V}_{i}= n \widehat{\Delta}_{n}-(n-1)\widehat{\Delta}_{n-1,i}; \qquad i=1,2,\cdots,n,
\end{equation}
where $\widehat{\Delta}_{n-1,i}$ is the estimator of $\Delta$ obtained from (\ref{est}) by using $(n-1)$ observations $X_1$, $X_2$, ..., $X_{i-1}$, $X_{i+1}$,..., $X_n;i=1,2,..,n$. The jackknife estimator $\widehat{\Delta}_{jack}$ of $\Delta$ is the average of the jackknife pseudo-values, that is
$$\widehat{\Delta}_{jack}=\frac{1}{n}\sum\limits_{i=1}^{n}\widehat{V}_{i}.$$
 The jackknife empirical likelihood of $\Delta$ is defined as
\begin{equation}\label{6}
 J(\Delta)=\sup_{\bf p} \left(\prod_{i=1}^{n}{p_i};\,\, \sum_{i=1}^{n}{p_i}=1;\,\,\sum_{i=1}^{n}{p_i (\widehat{V}_i}-\Delta)=0\right),
\end{equation}
 where ${\bf p}=(p_1,p_2,...,p_n)$ is a probability vector. The maximum of (\ref{6}) occurs at
\begin{equation*}
  p_i=\frac{1}{n}\left(1+\lambda(\widehat{V}_{i}-\Delta)\right)^{-1}, k=1,2,...,n,
\end{equation*}
where $\lambda$ is the solution of
\begin{equation}\label{eq7}
  \frac{1}{n}\sum_{i=1}^{n}{\frac{\widehat{V}_{i}-\Delta}{1+\lambda (\widehat{V}_{i}-\Delta)}}=0,
\end{equation}provided
\begin{equation*}
  \min_{{1\le i\le n}}\widehat{V}_{i}<\widehat \Delta<  \max_{1\le i\le n}\widehat{V}_{i}.
\end{equation*}
Also note that, $\prod\limits_{i=1}^{n}p_i$, subject to $\sum\limits_{i=1}^{n}p_i=1$, attains its maximum $n^{-n}$ at $p_i=n^{-1}$. Hence, the jackknife empirical log-likelihood ratio  for  $\Delta$ is given by
\begin{equation}\label{jelrat}
  l(\Delta)=-\sum_{i=1}^{n}\log\left[1+\lambda (\hat{V}_{i}-\Delta)\right].
\end{equation}

Next theorem explains the limiting distribution of $l(\Delta)$ which can be used to construct the JEL ratio  test for testing the independence between  $T$ and $J$. We are not providing a rigorous proof as it is direct from the Lemmas and Corollaries of Jing et al. (2009).
\begin{theorem}
  Suppose $E(\psi((T_1,J_1),(T_2,J_2)))^2<\infty$ and $\sigma^2>0$. Under $H_0$, as $n\rightarrow\infty$, $-2l(\Delta)$ converges in distribution to $\chi^{2}$ with one degree of freedom.
  \end{theorem}
   \par  In JEL ratio test, we reject $H_0$ in favor of $H_1$ if $$-2l(\Delta)> \chi^2_{1,\alpha},$$ where $\chi^2_{1,\alpha}$ is the upper $\alpha$ percentile point of $\chi^2$ distribution with one degree of freedom.
  
 \section{Simulation study}
 \par In this section, we conduct a Monte Carlo simulation study to evaluate the fine sample performance of our proposed JEL ratio test. For generating competing risks data with two causes, we consider the family of sub-distribution functions proposed by Dewan and Kulathinal (2007). Let
\begin{align}
\label{4.1}
F_1(t)=p_{1} F^{a}(t),\,\,
F_2(t)=F(t)-p_{1} F^{a}(t),
\end{align}
where $0\le p_{1} \le 0.5$, $1 \le a \le 2$ and $F(t)$ is a proper distribution function. The restrictions on parameters of the model are imposed due to the non-negativity condition of the cause-specific density function. It is clear that time to failure and cause of failure are independent when $a=1$ and $T$ and $J$ are dependent for all other values of $a$. Let $$F(t)=1-exp(-\lambda t), \,\,\lambda>0,\,t\ge 0.$$ be the distribution function employed in Eq.(\ref{4.1}) to generate lifetimes and corresponding causes. 
We simulated 10000 replications of random samples of sizes $n=20,40,60,80,100$ by considering different combinations of $(\lambda, p_1,a)$. As the results are similar, we report the results for parameter combinations $(\lambda, p_1)= (0.5,0.3)\text{and} (1,0.5)$ for various choices of $a$. For finding type I error, we set $a=1$ the results along with the type I error of the test statistic proposed by Dewan et al. (2004) is reported in Table 1. In Table 1, `JEL' represents the newly proposed test and `DDK' represents the test proposed by Dewan et al. (2004). It can be observed from Table 1 that the type I error of the proposed test converges to desired significance level as in case of Dewan et al. (2004).
\begin{table}[h]
\caption{Empirical type I error and power of the test compared with that of Dewan et al. (2004)}
\scalebox{1}{
\begin{tabular}{ccccc|cccccc}\hline
\multirow{2}{*}{} & \multicolumn{4}{c}{$(\lambda,p_1)=(0.5,0.3)$} & \multicolumn{4}{c}{$(\lambda,p_1)=(1,0.5)$}
\\ \cline{1-10}

\multirow{2}{*}{} & \multicolumn{2}{c}{JEL} & \multicolumn{2}{c}{DDK }&\multicolumn{2}{c}{JEL}&\multicolumn{2}{c}{DDK}
\\ \cline{1-10}

$n/\alpha$ &  $0.01$ &$0.05$&  $0.01$ &$0.05$ &$0.01$ &$0.05$&  $0.01$ &$0.05$& \\ \hline
\multirow{2}{*}{} & \multicolumn{8}{c}{$a=1$} & 
\\ \cline{1-10}
20&  0.0108& 0.0546 &0.0113 &0.0486& 0.0108 & 0.0513 & 0.0109 &0.0516 \\
40&  0.0106& 0.0534 &0.0108 &0.0491& 0.0094 & 0.0464 & 0.0088 &0.0485 \\
60&  0.0104& 0.0492 &0.0096 &0.0509& 0.0104 & 0.0474 & 0.0095 &0.0498  \\
80&  0.0104& 0.0492 &0.0096 &0.0509& 0.0104 & 0.0474 & 0.0095 &0.0498  \\
100& 0.0102& 0.0502 &0.0103 &0.0508& 0.0102 & 0.0486 & 0.0095 &0.0502 \\ \hline

\end{tabular}}
\end{table}

%
%
%
%
%
\par Here we  compare the  power of our test with the test proposed by Dewan et al. (2004) based on Kendall's tau.  Under $H_0$, their test statistic is asymptotically distributed as normal with mean zero and variance $\frac{4}{3}p_{1} (1-p_{1})$. We consider different choices for $a$ for illustration. The results of the power comparison are given in Table 2.

\begin{table}[h]
\caption{Empirical power of the test compared with that of Dewan et al. (2004)}
\scalebox{1}{
\begin{tabular}{ccccc|cccccc}\hline
\multirow{2}{*}{} & \multicolumn{4}{c}{$(\lambda,p_1)=(0.5,0.3)$} & \multicolumn{4}{c}{$(\lambda,p_1)=(1,0.5)$}
\\ \cline{1-10}

\multirow{2}{*}{} & \multicolumn{2}{c}{JEL} & \multicolumn{2}{c}{DDK }&\multicolumn{2}{c}{JEL}&\multicolumn{2}{c}{DDK}
\\ \cline{1-10}

$n/\alpha$ &  $0.01$ &$0.05$&  $0.01$ &$0.05$ &$0.01$ &$0.05$&  $0.01$ &$0.05$& \\ \hline
\multirow{2}{*}{} & \multicolumn{8}{c}{$a=1.3$} & 
\\ \cline{1-10}
20&  0.0108& 0.0546 &0.0113 &0.0486& 0.0108 & 0.0513 & 0.0109 &0.0516 \\
40&  0.0106& 0.0534 &0.0108 &0.0491& 0.0094 & 0.0464 & 0.0088 &0.0485 \\
60&  0.0104& 0.0492 &0.0096 &0.0509& 0.0104 & 0.0474 & 0.0095 &0.0498  \\
80&  0.0104& 0.0492 &0.0096 &0.0509& 0.0104 & 0.0474 & 0.0095 &0.0498  \\
100& 0.0102& 0.0502 &0.0103 &0.0508& 0.0102 & 0.0486 & 0.0095 &0.0502 \\ \hline
\multirow{2}{*}{} & \multicolumn{8}{c}{$a=1.5$} & 
\\ \cline{1-10}
20& 0.1546&	0.2134&	0.1679&	0.2236&	0.1782&	0.2490&0.1892&0.2576
  \\
40&0.2051&	0.4366&	0.1948&	0.4495&	0.2091&	0.4426&	0.2137&	0.4683
   \\
60& 0.2987&	0.5832&	0.2861&	0.5727&0.3445&	0.6414&	0.3681&	0.6339
   \\
80& 0.3875&	0.6102&	0.3984&	0.6268	&0.4449	&0.7106	&0.4694&	0.7624   \\
100& 0.4796&0.7531&	0.4733&	0.7592&	0.5465&	0.7716&	0.5606&	0.7963
 \\ \hline

\multirow{2}{*}{} & \multicolumn{8}{c}{$a=1.7$} & 
\\ \cline{1-10}
20& 0.2080&	0.2678&	0.1872&	0.2761&0.2087&	0.3281&	0.2198&	0.3562
  \\
40& 0.2456&	0.5321&	0.2153&	0.4942&	0.3241&	0.6168&	0.3058&	0.5900
  \\
60&0.3981&	0.6790&	0.3631	&0.6421&0.5484	&0.8242&	0.4971&	0.7782
    \\
80& 0.5028&	0.7901&	0.4694&	0.7447&	0.6752	&0.8904	&0.6436&	0.8674
   \\
100& 0.6109	&0.8623	&0.5883&	0.8119	&0.7976&	0.9201&	0.7390&	0.9035
 \\ \hline
\multirow{2}{*}{} & \multicolumn{8}{c}{$a=1.9$} & 
\\ \cline{1-10}
20&0.2394&	0.3482&	0.2292&	0.3297&0.2891&	0.4882&	0.2601&	0.4092
  \\
40& 0.3214&	0.6321&	0.2771&	0.5542&	0.4212&	0.7567&	0.3784&	0.6778&
  \\
60& 0.4561&	0.7902&	0.4234&	0.7382&0.6008&	0.8745	&0.5775&	0.8332
  \\
80& 0.6019&	0.8921&	0.5759&	0.8393&0.8071&	0.9263&	0.7464&	0.9136
  \\
100& 0.7494&	0.9532&	0.6909&	0.8924&	0.8786&	0.9827	&0.8390&	0.9551
 \\ \hline

\end{tabular}}
\end{table}
From Table 2, it is clear that our test has good power. As $a$ increases, the power of the test also increases. We can see that when $a=1.3$. the test proposed by Dewan et al. (2004) yields good power, but as $a$ increases our test performs more efficiently. The power of both tests increases with increase in  $a$ and sample size. This ensures the efficiency of the proposed method. 
\section{Data analysis}
The proposed testing procedure is applied to four real data sets for illustration. We use two data sets recently studied the in context of competing risks and two data sets which are exploited in competing risks analysis by other researchers, to validate our test.\\
\textit{Example 1}. First we use data obtained from a clinical trial on HIV infection and AIDS of 329 homosexual men from Amsterdam. The
data is available in R package `mstate' and is exclusively studied by Geskus (2015). During the course from HIV infection from non-syncytium-inducing (NSI) phenotype to death, intermediate events may occur that have an impact on subsequent disease progression. One such event is a switch of the HIV virus to the syncytium inducing (SI)
phenotype and the other is the progression to AIDS. Accordingly, the data consist of the lifetimes of 329 patients with corresponding cause of infection as SI or AIDS. Out of 329 patients for 114 patients`AIDS' was the first event to occur and for 108 patients the first event to occur was SI. The remaining 107 patients are observed to be event free in the study period and presently excluded from the analysis. The test statistic is obtained as $0.4827$. By comparing with chi square table values, we accept $H_0$ for this data at $5\%$ significance level and we conclude that time to failure and cause of failure are independent for this data. \\
\textit{Example 2}. Next, we consider the `fourD' data discussed in Beyersmann (2011). The data is available in the R-package `etm'. The 4D study was planned (Schulgen et al., 2005) and analysed (Wanner et al., 2005) for an event of interest in the presence of competing risks. The event of interest was defined as a composite of death from cardiac causes, stroke, and non-fatal myocardial infarction, whichever occurred first. The other competing event was death from other causes. Within the placebo group, there were 243 observed events of interest, 129 observed competing events,and 264 patients with censored event times. We consider the event of interest and other causes in our analysis.  For this data, we obtain the test statistic as $0.1491$. According we accept the null hypothesis of Independence of time to failure and cause of failure at 5\%  of significance. \\
\textit{Example 3}. Now, we illustrate the use of the proposed test using the data given  in Hoel (1972). The data contain the information about the survival time of mice, kept in a conventional germ free environment and all of which were  exposed to fixed dose of radiation at an age of 5 to 6 weeks. Lifetimes of 181 mice were observed, all of which are exactly observed lifetimes.  For each failure, the cause is either thymic lymphoma (cause 1), reticulam cell sacroma (cause 2) or other causes (cause 3). For each mouse we observed the exact failure and associated cause of failure. We combine the two types of cancer as a single cause while keeping the third cause as such.  The test statistic is estimated as $6.0764$. Now we accept $H_0$, that the time to failure and cause of failure are independent We can note that Dewan et al.(2004) also arrived at the same conclusion.\\
\textit{Example 4}. Next, we consider the failure time data from a laboratory test on pneumatic tires given in Davis and Lawrence (1989). This data set was given in Lawless (2011, Table 9.4). The test involved rotating the tires against a steel drum until some type of failure occurred. Failures were classified into six modes or categories: (i) open joint on the inner linear; (ii) rubber chunking on the shoulder; (iii) loose casing low on the side wall; (iv) cracking of the tread rubber; (v) cracking on the side wall; and (vi) any other causes.  Some of the tyres did not fail under the test, those lifetimes are not considered in this analysis. By observing the data mode of failure 4 seems to be the dominant cause of failure. So we combine all other modes pf failure into one and consider the data as a two risks problem as in Sankaran et al. (2010) in the context of comparing cumulative incidence functions. The test statistic value obtained is $20.2296$, which leads to the rejection of the null hypothesis. We can note that Sankaran et al. (2017) and Anjana et al. (2019) arrived at the same conclusion for this data. 
\vspace{-0.5cm}
\section{Concluding Remarks}
In this article, we developed a new testing procedure for independence of time to failure and cause of failure for a competing risks data. We use U-statistic theory and recently developed Jackknife empirical likelihood ratio test methodology to develop the test statistic. To the best of our knowledge, this is the first attempt to use JEL method in competing risks testing problem. The finite sample performance of the newly developed procedure is validated through a Monte Carlo simulation study which shows our test has better power than the test proposed by Dewan et al. (2004). Several data sets are used to illustrate the practical utility of the testing method. \\
The proposed test procedure does not incorporate right censored observations. JEL method to incorporate censoring observations are being studied. JEL ratio test for censored observations will be reported in a separate study. Further, we can develop JEL ratio test for the equality of cumulative incidence functions.
\section*{Acknowledgements}
We are thankful to Dr. Sudheesh K.K., ISI, Chennai for suggesting this research problem. 

\end{document}